\documentclass[aps,pra,twocolumn,showpacs,superscriptaddress,groupedaddress]{revtex4}  % for review and submission
\usepackage{graphicx}  % needed for figures
\usepackage{dcolumn}   % needed for some tables
\usepackage{bm}        % for math
\usepackage{amssymb}   % for math
\usepackage{graphicx}
\usepackage{amsmath}
\usepackage{inputenc}
\usepackage{verbatim}
\usepackage{color}
\usepackage{hyperref}
\usepackage{mathptmx}
\usepackage{amssymb}

\newcommand{\bra}[1]{\left\langle #1\right|}
\newcommand{\ket}[1]{\left| #1\right\rangle}
\newcommand{\braket}[2]{\left\langle
	#1\vphantom{#2}\right|\left.#2\vphantom{#1}\right\rangle}

\newcommand{\avg}[1]{\left\langle #1\right\rangle}
\newcommand{\be}[0]{\begin{equation}}
\newcommand{\ee}[0]{\end{equation}}
\newcommand{\intinf}[0]{\int\limits_{-\infty}^{+\infty}}

\newcommand{\lra}\simeq
\newcommand{\eeqref}[1]{Eq.~(\ref{#1})}

\newcommand{\pr}[0]{{\rm pr}}

\newcommand{\dd}[0] {{\rm d}}

\begin{document}

\title{Fisher information for far-field linear optical superresolution via homodyne or heterodyne detection in a higher-order local oscillator mode }
\author{Fan Yang}
\affiliation{Department of Physics and Astronomy, University of British Columbia, Vancouver, Canada, V6T 1Z4}
\author{Ranjith Nair}
\affiliation{Department of Electrical and Computer Engineering, National University of Singapore, 4 Engineering Drive 3, Singapore 117583}
\author{Mankei Tsang}
\affiliation{Department of Electrical and Computer Engineering, National University of Singapore, 4 Engineering Drive 3, Singapore 117583}
\affiliation{Department of Physics, National University of Singapore, 2 Science Drive 3, Singapore 117551}
\author{Christoph Simon}
\affiliation{Department of Physics and Astronomy and Institute for Quantum Science and Technology, University of Calgary, Calgary, Canada, T2N 1N4}
\author{Alexander I. Lvovsky}
\affiliation{Department of Physics and Astronomy and Institute for Quantum Science and Technology, University of Calgary, Calgary, Canada, T2N 1N4}
\affiliation{Russian Quantum Center, 100 Novaya St., Skolkovo, Moscow 143025, Russia}
\affiliation{Institute of Fundamental and Frontier Sciences, University of Electronic Science and Technology, Chengdu, Sichuan 610054, China}
\affiliation{P.N. Lebedev Physics Institute, Leninskiy Prospect 53, Moscow 119991, Russia}

\begin{abstract}
The distance between two point light sources is difficult to estimate if that distance is below the diffraction (Rayleigh's) resolution limit of the imaging device. A recently proposed technique enhances the precision of this estimation by exploiting the source-separation-dependent coupling of light into higher-order TEM modes, particularly the TEM$_{01}$ mode of the image. We theoretically analyze the estimation of the source separation by means of homodyne or heterodyne detection with a local oscillator in the TEM$_{01}$ mode, which is maximally sensitive to the separation in the sub-Rayleigh regime. We calculate the Fisher information associated with this estimation and compare it with direct imaging. For thermal sources, the Fisher information in any mode of the image plane depends nonlinearly on the average received photon number. We show that the per-photon Fisher information surpasses that of direct imaging (in the interesting sub-Rayleigh regime) when the average received  photon number per source exceeds two for homodyne detection and four for heterodyne detection.
\end{abstract}
\maketitle

\section{Introduction}

The resolution limit of conventional microscopes is determined by Rayleigh's criterion \cite{Rayleigh}. In the last few decades, various techniques have been invented to circumvent Rayleigh's limit by changing the imaging conditions. Such techniques utilize nonlinear optical properties of the object \cite{2Photon, STED}, near-field optics \cite{SNOM, durig} or work with photo-switchable samples \cite{STORM, PALM}. However, recently it was found that sub-Rayleigh resolution can be achieved for certain microscopy-related tasks  without resorting to nonlinear optics or near-field interactions. Such is the case, for example, for estimating the distance separating two point sources \cite{Tsang, Sliver, Ranjith, Lupo}.

The idea of this new approach was to count photons in the Hermite-Gaussian or transverse-electromagnetic (TEM) modes $\{\rm{TEM}_{0q}; q = 0,1, \ldots\}$ in the image plane \cite{Tsang}. The precision of this estimation has been calculated as the inverse of the Fisher information (FI) in accordance with the Cram\'er-Rao bound of classical statistics \cite{Bos, HarryL}. Remarkably, this FI is independent of the separation distance, in contrast to direct imaging in which the FI tends to zero in the limit of low separations. Moreover, it was shown that this method is quantum optimal, i.e. it permits extracting the maximum possible FI from each photon available to the observer \cite{Tsang}. Inspired by this analysis, a number of groups around the world demonstrated proof-of-principle experiments to achieve super-resolution \cite{Sheng, Tham, Paur, Yang}.

However, direct implementation of the scheme of Ref.~\cite{Tsang} requires a setup for spatial mode filtering in the Hermite-Gaussian basis, which is a challenge \cite{Lam2010,Vamivakas2016}. It is thus tempting to use a homodyne or heterodyne  detector instead of a mode filter, taking advantage of such a detector's sensitivity to the optical signal only in the mode that matches that of the local oscillator, which, in turn, can be readily prepared in any TEM mode by using spatial light modulators or optical cavities. Ref. \cite{Yang} demonstrated the viability of this method for achieving sub-Rayleigh resolution. The first-order mode $\rm{TEM}_{01}$ was chosen as the local oscillator in the experiments of Ref. \cite{Yang}.  As shown in Ref. \cite{Ranjith}, this mode contains most of the information on the source separation in the sub-Rayleigh regime. Consequently, we focus on dyne measurements of this mode in this paper.

Because homodyne and heterodyne detection are physically different from direct photon counting, the FI associated with these measurements needs to be evaluated independently. Ref.~\cite{tsanghomo} argues that homodyne detection offers no advantage with respect to direct imaging for weak thermal light because of the shot noise. However, there has been no similar analysis for arbitrary thermal sources. Here we show that homodyne and heterodyne detection do possess an advantage over direct imaging for estimating separations well below the Rayleigh limit when the average received photon number per-source of the thermal state exceeds two and four respectively.

\section{Concepts}
\subsection{Displacement and TEM$_{01}$ mode}
To illustrate the mode transformation and detection process, we begin with a brief description of homodyne detection in TEM$_{01}$ using classical optics. A complete quantum optical derivation that includes the effect of shot noise is given in the later sections and appendices. Heterodyne detection is closely related, see below. We work in a single transverse dimension and assume quasi-monochromatic light in the paraxial approximation. We also assume a translationally invariant imaging system with a Gaussian point spread function. With such assumptions, a pointlike light source located at the optical axis of the objective lens is imaged in the TEM$_{00}$ mode. When the light source is displaced by $\pm d$, the beam amplitude in the image plane is
\begin{equation}
\alpha E_0(x\pm d)=\alpha\left(\frac{1}{2 \pi \sigma^2}\right)^{1/4}e^{-\left(\frac{x\pm d}{2\sigma}\right)^2},\label{profile}
\end{equation}
where $\alpha$ is the amplitude, $E_0(x)$ is the normalized amplitude profile of TEM$_{00}$ and $\sigma$ is the beam width.
For small displacement, this can be approximated by Taylor expansion
\begin{equation}
\alpha E_0(x\pm d)\approx\alpha E_0(x)\pm\alpha d\cdot E_0'(x)=\alpha E_0(x)\mp \frac{d}{2\sigma}\alpha E_1(x),\label{amplitude}
\end{equation}
where $E_0'(x)$ is the derivative of $E_0(x)$ with respect to $x$ and $E_1(x)$ is the normalized amplitude profile of TEM$_{01}$. This means that, when the source becomes displaced, TEM$_{01}$ acquires a nonzero amplitude that is $\mp \frac d{2\sigma}$ of the amplitude of TEM$_{00}$, and a nonzero power corresponding to $d^2/(4\sigma^2)$ of that in TEM$_{00}$. We detect the image by homodyne detection with the local oscillator prepared in TEM$_{01}$, resulting in a photocurrent proportional to the displacement $d$ (with added shot noise, which is included later in our quantum formalism of analysis). A null measurement of the displacement is thereby achieved,  in contrast to direct imaging, in which a signal in the form of a certain intensity distribution is present for all displacements.

\subsection{Fisher information (FI)}
As is standard in astronomy \cite{EricD, JonasZ}, single-molecule microscopy \cite{JerryChao},  asymptotic statistics \cite{AW}, and engineering statistics \cite{HarryL}, we adopt here the FI as the sensitivity measure. Given a probability distribution pr$(Y|\lambda)$ of measurement outcome $Y$ as a function of parameter $\lambda$, the FI is defined as
\begin{equation}
F_\lambda=\left<\left(\frac\partial{\partial\lambda}\log \text{pr}(Y|\lambda)\right)^2\right>
\end{equation}
where $\avg\cdot$ represents statistical average.

The inverse of FI gives the Cram\'er-Rao bound, which is a lower bound on the mean-square error of any unbiased estimator. The bound can be attained in the asymptotic limit of infinite repetitions by the maximum-likelihood estimator \cite{AW,HarryL}. Although a biased estimator can violate the Cram\'er-Rao bound for limited repetitions \cite{Tham,MT}, one can generalize the bound for any biased or unbiased estimator by adopting a modified error criterion from the Bayesian or minimax perspective \cite{MT}. In particular, the Bayesian Cram\'er-Rao bound by Sch\"utzenberger \cite{MP} and Van Trees \cite{HarryL,MT,RichardD} and the local asymptotic minimax theorem by H\'ajek and Le Cam \cite{AW,RichardD} are valid for any biased or unbiased estimator, and both depend on the FI.

For the imaging problem, the probability distribution and therefore the FI depend on the optical measurement method. In this paper, we compare the FI of three measurements---homodyne detection, heterodyne detection and direct imaging---of the light on the image plane from two thermal point sources. Let us note that the quantum Fisher information computed in Refs.\cite{Tsang,Ranjith,Lupo} is an upper bound on the FI for any measurement allowed by quantum mechanics, but otherwise outside the scope of this paper.

\section{Measuring the displacement of a single source}
\subsection{Coherent source}
The noise properties of homodyne and heterodyne detection of coherent or thermal fields can be discussed using either a semiclassical or quantum-optical formalism. Since both approaches give exactly the same quantitative results\cite{Shapiro}, the choice of formalism is a matter of taste and familiarity. Here, we use the quantum-optical formalism to make explicitly sure that our measurement models agree with quantum mechanics.

In order to introduce our approach for calculating the per-photon FI, we first consider a single coherent source. As is evident from Eq. (\ref{amplitude}), a coherent state $\ket{\alpha}$ in TEM$_{00}$ displaced by $\pm d$ is approximately equivalent to the direct product
\begin{equation}\label{alphapm}
\ket{\alpha}_\pm=\ket{\alpha}_0\otimes\ket{\mp\frac{d}{2\sigma}\alpha}_1,
\end{equation}
where the subscripts 0 and 1 label the two lowest order TEM modes centered on the optical axis of the lens. A full quantum optical analysis leading to \eeqref{alphapm} is given in Appendix A. Importantly, displacements in opposite directions give rise to opposite amplitudes of the TEM$_{01}$ component because of the antisymmetric shape of that mode.

Without loss of generality, we assume $\alpha$ to be real. The homodyne detector will measure the probability distribution of the quadrature $X$ in the state $\ket{\mp\frac{d}{2\sigma}\alpha}$, which is given by \cite{Leonhardt}
\begin{equation}
\pr_\alpha(X|d)=\bigg|\bigg<X\bigg|\mp\frac{d}{2\sigma}\alpha\bigg>\bigg|^2=\frac{1}{\sqrt\pi}\text{exp}\Bigg[{-\left(X\pm\frac{d}{2\sigma}\sqrt{2}\alpha\right)^2}\Bigg].
\end{equation}
A single quadrature measurement yields  a sample of this distribution, from which the displacement $d$ can be estimated. The FI is given by
\begin{equation}
\begin{split}
F^{(\alpha)}(d)&=\left<\left(\frac{\partial}{\partial d} \log{\pr_\alpha(X|d)}\right)^2\right>
\\&=\intinf\left(\frac{\partial}{\partial d} \log{\pr_\alpha(X|d)}\right)^2\pr_\alpha(X|d)\dd X=\frac{\alpha^2}{\sigma^2}.
\end{split}
\end{equation}
For the coherent state, the average photon number $N=\alpha^2$, so the per-photon FI is $F^{(\alpha)}_1(d)=1/\sigma^2$. We notice that Ref. \cite{Hsu} also analyzes the performance of homodyne detection with TEM$_{01}$ mode to estimate the displacement of a single coherent light source by calculating the quantum noise limited sensitivity.

\subsection{Thermal source}

Let us now consider a single thermal source. If the average photon number of the original thermal state is $N$, then we have a thermal state with average photon number $\frac{d^2}{4\sigma^2}N$ in TEM$_{01}$. This follows from the fact that linear mode transformations (beam splitters) map thermal states into (in general correlated) states, each single mode of which is in a thermal state. It can also be verified more formally using the Sudarshan-Glauber $P$-representation, as is shown in Appendix B.

A thermal state with average photon number $N$ can be described by the Wigner function \cite{Leonhardt}
\begin{equation}
W(X,P)=\frac{1}{\pi (2N+1)}\exp\left[-\frac{X^2+P^2}{2N+1}\right].
\end{equation}
For a single thermal light source displaced by $d$, the Wigner function of the thermal state in TEM$_{01}$ is therefore
\begin{equation}
W^{(1)}_{01}(X,P)=\frac{1}{\pi \left(\frac{d^2}{2\sigma^2}N+1\right)}\text{exp}\left[-{\frac{X^2+P^2}{\frac{d^2}{2\sigma^2}N+1}}\right].
\end{equation}
Using homodyne detection, we obtain the distribution of quadrature $X$ in TEM$_{01}$\cite{Leonhardt}
\begin{equation}
\pr_{01}^{(1)}(X|d)=\frac{1}{\sqrt{\pi\left(\frac{d^2}{2\sigma^2}N+1\right)} }\exp\left[-\frac{X^2}{\frac{d^2}{2\sigma^2}N+1}\right].
\end{equation}
The width of this Gaussian distribution depends on $d$, and hence a single sample thereof permits inferring this parameter. We find the FI for this inference to be
\begin{equation}
F^{(1)}_{N}(d)=\left<\left(\frac{\partial}{\partial d} \log{\pr_{01}^{(1)}(X|d)}\right)^2\right>=\frac{2d^2N^2}{(d^2 N+2\sigma^2)^2},
\label{homFI}
\end{equation}
and the per-photon FI is
\begin{equation}
F^{(1)}_{1}(d)=\frac{1}{N} F^{(1)}_{N}(d)=\frac{2d^2N}{(d^2 N+2\sigma^2)^2}.
\end{equation}
An important observation we can make here is that $F^{(1)}_1(d)$ depends on $N$, i.e.~the FI is not additive with respect to the number of incoming photons. For example, for low $N$, $F^{(1)}_1(d)\approx d^2N/2\sigma^4$, which means that performing a single measurement of $d$ on a mode with $N$ photons gives a higher precision than two separate measurements on a mode with $N/2$ photons.

In practice it is often advantageous to use heterodyne rather than homodyne detection (i.e. a local oscillator with a slightly different frequency) in order to reduce flicker noise \cite{Yang}. Heterodyne detection is formally equivalent to mixing the input light with vacuum on a 50/50 beam splitter, followed by a homodyne detection of orthogonal quadratures in the two output modes \cite{Wiseman}. The overall FI for heterodyne detection can be obtained from that for homodyne detection by the following steps: replace $N$ by $N/2$, and multiply by 2 to take into account the fact that two quadratures with independent statistics are measured. The per-photon FI is still obtained by dividing by $N$. Since the numerator in Eq. (\ref{homFI}) is quadratic in $N$, this means that the per-photon FI for heterodyne detection can be obtained from that for homodyne detection simply by replacing $N$ with $N/2$.

\section{Measuring the separation of two thermal sources}

\subsection{Homodyne and heterodyne detection}

In practice, we are most interested in measuring the distance between two point light sources (e.g. stars) separated below the Rayleigh limit, for which the direct imaging approach offers reduced precision \cite{Tsang}. For two thermal sources each with average photon number $N$ and displaced by $\pm d$, the average photon number detected in TEM$_{01}$ is the sum of the photon numbers from each state, because the random phase between the sources does not lead to intereference when we sum up the photon numbers. From the results of Sec. III.B, the $\rm{TEM}_{01}$ mode is in a thermal state of average photon number $\frac{d^2}{2\sigma^2}N$.

We can then perform the same calculation as in the previous section. In this case, the Wigner function for the light in TEM$_{01}$ is
\begin{equation}
W^{(2)}_{01}(X,P)=\frac{1}{\pi \left(\frac{\theta^2}{4\sigma^2}N+1\right)}\text{exp}\left[{-(X^2+P^2)\bigg/\left(\frac{\theta^2}{4\sigma^2}N+1\right)}\right],
\end{equation}
where  we choose to work with $\theta=2 d$, the separation of the light sources. This corresponds to the distribution of the $X$ quadrature
\begin{equation}\label{pr2S}
\pr_{01}^{(2)}(X|\theta)=\frac{1}{\sqrt{\pi\left(\frac{\theta^2}{4\sigma^2}N+1\right)} }\text{exp}\left[{-X^2\bigg/\left(\frac{\theta^2}{4\sigma^2}N+1\right)}\right],
\end{equation}
and the per-photon FI
\begin{equation}\label{F12S}
F^{(2)}_1(\theta)=\frac{\theta ^2 N }{(\theta^2 N+4\sigma^2)^2}.
\end{equation}
Following the same arguments as in the previous section, the per-photon FI for heterodyne detection can be obtained by replacing $N$ with $N/2$ in Eq. (\ref{F12S}).

As a side remark, since none of the calculations depend on the two sources being of equal strength, the total FI under either detection method for two sources of unequal strengths $N_1$ and $N_2$ can be obtained by replacing $N$ with $(N_1+N_2)/2$.

\subsection{Direct imaging}

We now evaluate the FI for spatially-resolved direct imaging of two incoherent thermal sources. An exact expression for this quantity for arbitrary source strengths is unknown and appears to be difficult to obtain.  We can, however, approximate it by noting that, in practice, the photon counts on each pixel are integrated over a large number of temporal modes, and their statistics can be approximated as Gaussian by virtue of the central limit theorem. The FI then becomes simple to evaluate, as shown in Appendix C. We also find that this approximate FI for any $N$, when evaluated on the per-photon basis, is upper-bounded by the per-photon FI in the $N\ll 1$ limit. Thus we simply show the calculation of the FI for $N\ll 1$ here and use it as an upper bound on the approximate FI for arbitrary $N$.

In the $N\ll 1$ limit, information comes from one-photon events only, and the per-photon FI can be computed from the probability distribution of each photon. For direct imaging, the measurement outcome is the position of arrival $x$ of the photon in the image plane whose probability density is
\begin{equation}\label{prDI}
\pr^{(DI)}(x|\theta)=\frac{1}{2\sqrt{2\pi} \sigma}\left[e^{-\frac{(x-\theta/2)^2}{2\sigma^2}}+e^{-\frac{(x+\theta/2)^2}{2\sigma^2}}\right]
\end{equation}
The per-photon information is hence
\begin{equation}
\begin{split}
F_1^{(DI)}(\theta)
&=\int_{-\infty}^\infty \dd x\, \bigg[\frac{(2x-2xe^{\frac{x\theta}{\sigma^2}}+\theta+\theta e^{\frac{x\theta}{\sigma^2}})^2}{32\sqrt{2\pi}(1+e^{\frac{x\theta}{\sigma^2}})\sigma^5}e^{-\frac{(x+\theta/2)^2}{2\sigma^2}}\bigg]\\
&=\frac1{4\sigma^2}-\frac1{2\sqrt{2\pi}\sigma^5}\int_{-\infty}^\infty\dd x\,\frac{x^2 e^{-(x+\theta/2)^2/2\sigma^2}}{1+e^{-x\theta/\sigma^2}}
\end{split}
\label{directintegral}
\end{equation}
which can be evaluated numerically.

\begin{figure}\label{compgraph}
\centering
\includegraphics[width=\columnwidth]{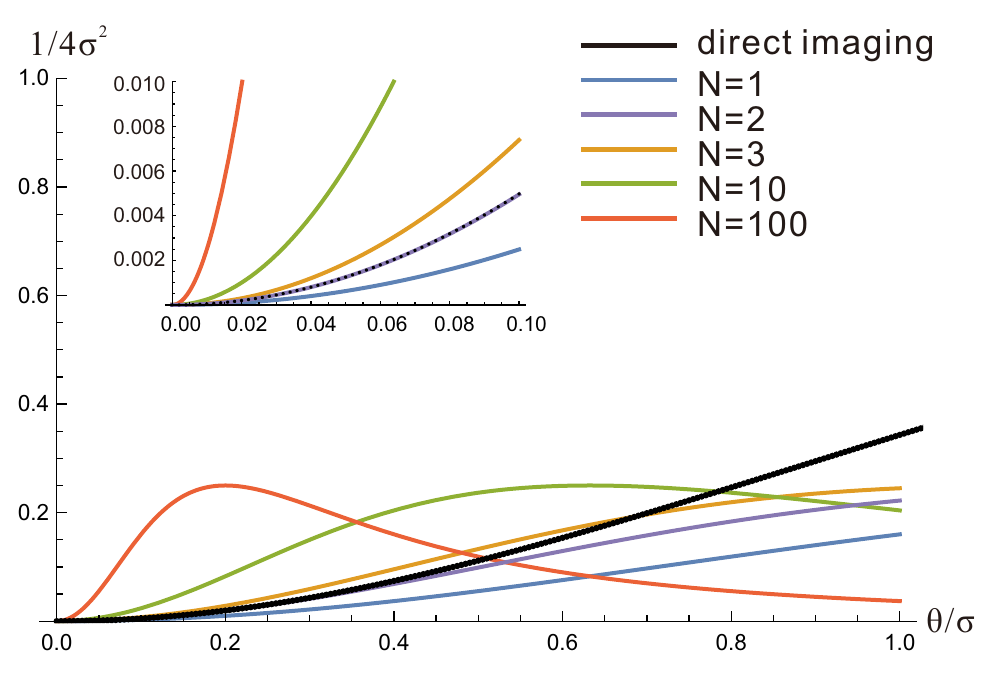}
\caption{Per-photon FI of homodyne and direct imaging for two incoherent thermal sources. The (black) direct imaging curve is for weak thermal states with $N\ll1$, which gives the approximate FI for direct imaging (see main text and appendix C).  Inset: per-photon FI for $\theta\ll\sigma$. In this limit, the FI for direct imaging is the same as homodyne detection when $N=2$. Black dots in the inset represent the result for direct imaging. Note that when $\theta/\sigma$ becomes close to 1, the population in TEM$_{01}$ starts to decrease because higher-order modes become important. This higher-order mode effect is not included in the figure. Our results for the FI are accurate in the regime of small $\theta/\sigma$. The per-photon FI for heterodyne detection can be obtained from that for homodyne detection by replacing $N$ with $N/2$, see text.}
\end{figure}

\subsection{Comparison}

In Fig.~1, we plot the per-photon FI for direct imaging and homodyne detection. One sees that homodyne detection is advantageous for small separations as long as the average photon number $N>2$, which is consistent with the conclusion in Ref. \cite{tsanghomo} that there is no advantage for small $N$. This also means that heterodyne detection is advantageous for $N>4$. This advantage can be understood by noting that, for small separations $\theta\ll\sigma$, the FI in Eq. (\ref{F12S}) for homodyne detection scales as $\theta^2 N/16\sigma^4$, whereas for direct imaging it scales like $\theta^2/8\sigma^4$ [see Eq.~(\ref{directintegral}) and Appendix C].

The maximum per-photon FI with homodyne detection is achieved for $\theta^2 N=4\sigma^2$ (i.e. when  there are $\frac{d^2}{4\sigma^2}N=\frac 14$ photons in TEM$_{01}$ mode per source) and equals $1/16\sigma^2$. This corresponds to $1/4$ of the per-photon FI obtained in the quantum optimal measurement, which is achieved by means of a photon number measurement in TEM$_{01}$ \cite{Tsang}.

Let us compare the three methods (direct imaging, homodyne detection in TEM$_{01}$ and  photon number measurement in TEM$_{01}$) in order to better understand the difference in their precision. For a direct image, the probability distribution Eq. (\ref{prDI}) of a photon's position of arrival in the image plane is a sum of two Gaussians, which, for $\theta\ll\sigma$, are almost indistinguishable from a single Gaussian centered $x=0$, and hence the inference on the source separation $\theta$ is very poor. The photon number measurement, on the other hand, is a null measurement: the signal power is proportional to $\theta^2$, so there is no signal whatsoever at $\theta=0$,  leading to a FI that is independent of $\theta$. The homodyne measurement lies in between. It is not an ideal null measurement because of the shot noise, but it can still give a substantial advantage relative to direct imaging for sufficiently large $N$. Eq. (\ref{F12S}) shows that maximum sensitivity for the homodyne measurement is achieved when $N \theta^2=4\sigma^2$.

The requirement of $N>2$ (or $N>4$) means that the homodyne (or heterodyne) approach is most promising for measurements of distances between objects with rough surfaces that scatter laser light. Such measurements can occur, for example, in LIDARs that are used to operate autonomous vehicles. Since the laser can have a very large number of photons in a single mode, the scattered light is likely to contain speckles with multiple photons per mode. For astronomical applications, e.g.~measurements of distances between binary star components, the advantages and disadvantages of this method require further analysis to account not only for fundamental noise sources, but also for technical issues, such as atmospheric turbulences.

\setcounter{figure}{0}

\setcounter{equation}{0}

\appendix

\section{Coherent light}
Here we prove \eeqref{alphapm}. Coherent states can be generated by the phase-space displacement operator $D[\alpha;a_k^{\dagger}]=e^{-|\alpha|^2/2}e^{\alpha a_k^{\dagger}}$ acting on the vacuum state $\ket{\rm vac}$, where $a_k^{\dagger}$ is the creation operator for mode $k$. We can define the creation operators for the Hermite-Gaussian modes TEM$_{00}$ and TEM$_{01}$ as $a_{0,1}^\dagger=\int\dd x\, E_{0,1}(x)a_x^\dagger$ with the corresponding subscripts. Then, the creation operator for a beam that is physically displaced by $\pm d$ is $a_{\pm}^\dagger=\int\dd x\, E_0(x\pm d)a_x^\dagger$, where $a_x^\dagger$ is the creation operator at position $x$ in the image plane.

From \eeqref{amplitude}, we have, to leading order,
\begin{equation}\label{a}
a_\pm^\dagger= a_0^\dagger\mp\frac{d}{2\sigma}a_1^\dagger
\end{equation}
For a single coherent light source in TEM$_{00}$, the states corresponding to displacements by $\pm d$ can be expressed as
\begin{equation}
\ket{\alpha}_{\pm}=D[\alpha;a_{\pm}^{\dagger}]\ket{\rm vac}=e^{-|\alpha|^2/2}e^{\alpha a_\pm^\dagger}\ket{\rm vac}
\end{equation}
Using Eq. (\ref{a}), we have
\begin{equation}
\begin{split}
\ket{\alpha}_\pm&=e^{|\frac{d}{2\sigma}\alpha|^2/2}D[\alpha; a_0^{\dagger}]D\left[\mp\frac{d}{2\sigma}\alpha; a_1^\dagger\right]\ket{\rm vac}\\
&=\ket{\alpha}_0\otimes\ket{\mp\frac{d}{2\sigma}\alpha}_1,
\end{split}
\end{equation}
The prefactor $e^{|\frac{d}{2\sigma}\alpha|^2/2}\simeq 1$ for sufficiently small $d/\sigma$.

\section{Incoherent light}
We now obtain a similar result for thermal sources: if the physically displaced source is in a thermal state with the average photon number $N$, then TEM$_{01}$ will contain a thermal state with average photon number $\frac{d^2}{4\sigma^2}N$. For a single incoherent thermal light source, we write the density matrix using the Sudarshan-Glauber $P$-representation,
\begin{equation}
\rho=\int P(\alpha)\ket{\alpha}_\pm\bra\alpha\, \dd^2 \alpha,
\end{equation}
where $P(\alpha)=\frac{1}{\pi N}e^{-|\alpha|^2/N}$ is the $P$ function of the thermal state with $N$ photons.
We substitute \eeqref{alphapm} and take the partial trace of $\rho$ over TEM$_{00}$ to get the density matrix $\rho_1$ in TEM$_{01}$
\begin{equation}\label{sup1}
\begin{split}
\rho_1&=\text{Tr}_0\rho=\frac{1}{\pi}\int\,_0\bra{\beta} \rho \ket{\beta}_0\dd^2\beta\\
&=\frac{1}{\pi^2 N}\int\dd^2\alpha\, e^{|\alpha|^2/N}\int\dd^2\beta\, _0\bra{\beta}\left(\ket{\alpha}_\pm\bra{\alpha}\right)\ket{\beta}_0\\
&=\frac{1}{\pi^2 N}\int\dd^2\alpha\bigg\{e^{|\alpha|^2/N}\, \ket{\mp\frac{d}{2\sigma}\alpha}_1\bra{\mp\frac{d}{2\sigma}\alpha}\\
&\hspace{0.4cm}\times\int\dd^2\beta\, |_0\braket{\beta}{\alpha}_0|^2\bigg\}
\end{split}
\end{equation}
Notice that
\begin{equation}
\int\dd^2\beta\, |_0\braket{\beta}{\alpha}_0|^2={_0\bra{\alpha}}\left[\int\dd^2\beta\, \ket{\beta}_0\bra{\beta}\right]\ket{\alpha}_0=\pi,
\end{equation}
where we utilize the fact that
\begin{equation}
\int\ket\beta\bra\beta\dd^2\beta=\pi.
\end{equation}
$\rho_1$ now reads
\begin{equation}
\rho_1=\frac{1}{\pi N}\int\dd^2\alpha\, e^{|\alpha|^2/N}\, \ket{\mp\frac{d}{2\sigma}\alpha}_1\bra{\mp\frac{d}{2\sigma}\alpha}.
\end{equation}
At last, we change the integration variable and obtain
\begin{equation}\label{rho1}
\rho_1=\frac{1}{\pi N_1}\int e^{-|\alpha|^2/N_1}\ket{\alpha}_1\bra\alpha \dd^2\alpha
\end{equation}
where $N_1=\frac{d^2}{4\sigma^2}N$ is the average photon number in TEM$_{01}$. $\rho_1$ describes a thermal state with average photon $N_1$.

\section{Direct imaging}
Let $n^{(m)} = (n_1^{(m)},n_2^{(m)},\dots)^\top$ be a column vector of photon counts in the $m$th temporal mode, where each $n_j^{(m)}$ is the photon count in a spatial mode and $\top$ denotes the transpose. Integrated over $M$ temporal modes, the photon-count vector is $n = \sum_{m=1}^M n^{(m)}$. Assuming a large $M$ and independent and identically distributed statistics across the temporal modes, the statistics of $n$ can be approximated as Gaussian by virtue of the central limit theorem. Defining $\mu \equiv \langle n^{(m)}\rangle$ and $\Sigma \equiv \langle n^{(m)}n^{(m)\top}\rangle-\langle n^{(m)}\rangle\langle n^{(m)}\rangle^\top$, the mean of $n$ is $M\mu$ and the covariance matrix is $M\Sigma$. The per-photon FI for
the normally distributed $n$ becomes
\cite{Kay}
\begin{align}
F^{(DI)}_1(\theta) &\approx
\frac{1}{N}
\frac{\partial \mu^\top}{\partial \theta}\Sigma^{-1}\frac{\partial\mu}{\partial\theta}
\nonumber\\&\quad
+ \frac{1}{2MN} \operatorname{tr}\left(\Sigma^{-1}
\frac{\partial\Sigma}{\partial\theta}
\Sigma^{-1}\frac{\partial\Sigma}{\partial\theta} \right),
\\
\lim_{M\to\infty}F^{(DI)}_1(\theta)
&= \frac{1}{N}
\frac{\partial \mu^\top}{\partial \theta}\Sigma^{-1}\frac{\partial\mu}{\partial\theta}.
\end{align}
This quantity is upper-bounded by the per-photon FI in the $N \ll 1$ limit; the proof is as follows. Define the normalized mean count vector as $p \equiv \mu/N$. Using the optical equivalence theorem \cite{Mandel}, it can be shown that $\Sigma = N D + V$, where $D_{jk} = p_{j}\delta_{jk}$, $V_{jk} = E(|\alpha_j|^2|\alpha_k|^2)-E(|\alpha_j|^2)E(|\alpha_k|^2)$, $\alpha_j$ denotes the c-number amplitude of each spatial mode, and $E$ denotes the expectation with respect to the $P$ function. Since the $P$ function is classical, $V$ is a covariance matrix and must be positive-semidefinite, resulting in the matrix inequalities $\Sigma \ge ND$ and $\Sigma^{-1} \le (ND)^{-1}$. Hence
\begin{align}
\frac{1}{N}\frac{\partial \mu^\top}{\partial \theta}\Sigma^{-1}
\frac{\partial\mu}{\partial\theta} &\le
\frac{\partial p^\top}{\partial \theta}D^{-1}
\frac{\partial p}{\partial\theta}
=\sum_j \frac{1}{p_{j}}\left(\frac{\partial p_{j}}{\partial \theta}\right)^2,
\end{align}
the last expression of which does not
depend on $N$ and coincides with the per-photon FI in the $N\ll 1$ limit.

For $N\ll1$, we find the per-photon FI for direct imaging at small separation $\theta$ using
\begin{equation}
F_1^{(DI)}(\theta)=\int_{-\infty}^\infty\frac{\left[\partial\pr^{(DI)}(x|\theta)/{\partial\theta}\right]^2}{\pr^{(DI)}(x|\theta)}\dd x
\end{equation}
where $\pr^{(DI)}(x|\theta)$ is given by \eeqref{prDI}. To the leading order in $\theta$, we have
$$\frac\partial{\partial\theta}\pr^{(DI)}(x|\theta)\approx-\frac{\theta}{4\sqrt{2\pi} \sigma^3}\left(1-\frac{x^2}{\sigma^2}\right)e^{-\frac{x^2}{2\sigma^2}}$$
and
$$\pr^{(DI)}(x|\theta)\approx\frac{1}{\sqrt{2\pi} \sigma}e^{-\frac{x^2}{2\sigma^2}},
$$ hence
\begin{equation}
F_1^{(DI)}(\theta)=\frac{\theta^2}{8\sigma^4}+O\left(\theta^3\right).
\end{equation}

\begin{acknowledgments}
F.Y., C.S. and A.L. acknowledge financial support from NSERC. R.N. and M.T. acknowledge support from the Singapore Ministry of Education Academic Research Fund Tier 1 Project R-263-000-C06-112. F.Y. and A.L. acknowledge financial support from CIFAR.
\end{acknowledgments}

\end{document}